\documentclass[a4paper,12pt]{article}
\usepackage{amsthm,amsmath,amssymb,latexsym,bm}

\newtheorem{thm}{Theorem}[section]

\newtheorem{rem}[thm]{Remark}
\numberwithin{equation}{section}

\newcommand{\ds}{\displaystyle}
\newcommand{\vep}{\varepsilon}

\newcommand{\wh}{\widehat}
\newcommand{\mf}{\mathfrak}

\newcommand{\DSPainleveKnown}{
\begin{table}
\[\begin{array}{c@{\qquad}c@{\qquad}c}\hline
	\text{Lie algebra}& \text{Partition}&
	\text{Painlev\'{e} system}\\[4pt]\hline
	A_1^{(1)}& (2)& P_{\rm{II}}\\[4pt]
	& (1,1)& P_{\rm{IV}}\\[4pt]
	A_2^{(1)}& (3)& P_{\rm{IV}}\\[4pt]
	& (2,1)& P_{\rm{V}}\\[4pt]
	& (1,1,1)& P_{\rm{VI}}\\[4pt]
	A_3^{(1)}& (4)& P_{\rm{V}}\\[4pt]
	A_4^{(1)}& (5)& \mathcal{H}^{A_4^{(1)}}\\[4pt]
	A_5^{(1)}& (6)& \mathcal{H}^{A_5^{(1)}}\\[4pt]\hline
\end{array}\]
\caption{Relation between $A_n^{(1)}$-hierarchies and Painlev\'{e} systems}
\end{table}
}

\newcommand{\DSPainleveResult}{
\begin{table}
\[\begin{array}{c@{\qquad}c@{\qquad}c}\hline
	\text{Lie algebra}& \text{Partition}&
	\text{Painlev\'{e} system}\\[4pt]\hline
	A_3^{(1)}& (2,2)& P_{\rm{VI}}\\[4pt]
	& (3,1)& \mathcal{H}^{A_4^{(1)}}\\[4pt]
	A_4^{(1)}& (4,1)& \mathcal{H}^{A_5^{(1)}}\\[4pt]
	& (2,2,1)& \text{system \eqref{Eq:CP6} with \eqref{Eq:CP6_Ham}}\\[4pt]
	A_5^{(1)}& (3,3)&
	\text{system \eqref{Eq:CP6} with \eqref{Eq:CP6_Ham}}\\[4pt]\hline
\end{array}\]
\caption{List of Painlev\'{e} systems obtained in this article}
\end{table}
}

\title{Drinfeld-Sokolov hierarchies of type $A$ and fourth order Painlev\'{e} systems}
\author{Kenta Fuji and Takao Suzuki\\
{\small Department of Mathematics, Kobe University}\\
{\small Rokko, Kobe 657-8501, Japan}\\
{\small E-mail: suzukit@math.kobe-u.ac.jp}}
\date{}

\begin{document}

\maketitle

\begin{abstract}
We study the Drinfeld-Sokolov hierarchies of type $A_n^{(1)}$ associated with the regular conjugacy classes of $W(A_n)$.
A class of fourth order Painlev\'{e} systems is derived from them by similarity reductions.
\end{abstract}

\section{Introduction}

Three types of fourth order Painlev\'{e} type ordinary differential equations have been studied \cite{FS,NY1,S}.
They are extensions of the Painlev\'{e} equations $P_{\rm{II}},\ldots,P_{\rm{VI}}$ and expressed as Hamiltonian systems
\[
	\mathcal{H}^{X_n^{(1)}}:\quad
	\frac{dq_i}{dt} = \frac{\partial H^{X_n^{(1)}}}{\partial p_i},\quad
	\frac{dp_i}{dt} = -\frac{\partial H^{X_n^{(1)}}}{\partial q_i}\quad
	(i=1,2),
\]
with the Coupled Hamiltonians
\[\begin{split}
	H^{A_4^{(1)}} &= H_{\rm{IV}}(q_1,p_1;\alpha_2,\alpha_1)
	+ H_{\rm{IV}}(q_2,p_2;\alpha_4,\alpha_1+\alpha_3) + 2q_1p_1p_2,\\
	tH^{A_5^{(1)}} &= H_{\rm{V}}(q_1,p_1;\alpha_2,\alpha_1,\alpha_1+\alpha_3)\\
	&\quad + H_{\rm{V}}(q_2,p_2;\alpha_4,\alpha_1+\alpha_3,\alpha_1+\alpha_3)
	+ 2q_1p_1(q_2-1)p_2,\\
	t(t-1)H^{D_6^{(1)}} &= H_{\rm{VI}}(q_1,p_1;\alpha_0,\alpha_3+\alpha_5,
	\alpha_3+\alpha_6,\alpha_2(\alpha_1+\alpha_2))\\
	&\quad + H_{\rm{VI}}(q_2,p_2;\alpha_0+\alpha_3,\alpha_5,\alpha_6,
	\alpha_4(\alpha_1+2\alpha_2+\alpha_3+\alpha_4))\\
	&\quad + 2(q_1-t)p_1q_2\{(q_2-1)p_2+\alpha_4\},
\end{split}\]
where
\[\begin{split}
	H_{\rm{IV}}(q,p;a,b) &= qp(p-q-t) - aq - bp,\\
	H_{\rm{V}}(q,p;a,b,c) &= q(q-1)p(p+t) + atq + bp - cqp,\\
	H_{\rm{VI}}(q,p;a,b,c,d) &= q(q-1)(q-t)p^2 - \{(a-1)q(q-1)\\
	&\quad +bq(q-t)+c(q-1)(q-t)\}p + dq.
\end{split}\]
But complete classification of fourth order Painlev\'{e} systems is not achieved, so that the existence of unknown ones is expected.
In this article, we derive a class of fourth order Painlev\'{e} systems from the Drinfeld-Sokolov hierarchies of type $A_n^{(1)}$ by similarity reductions.

The Drinfeld-Sokolov hierarchies are extensions of the KdV (or mKdV) hierarchy for the affine Lie algebras \cite{DS}.
For type $A_n^{(1)}$, they imply several Painlev\'{e} systems by similarity reductions \cite{AS,KIK,KK1,KK2,NY1}; {\it see Table 1}.\DSPainleveKnown
Such fact clarifies the origines of several properties of the Painlev\'{e} systems, Lax pairs, affine Weyl group symmetries and particular solutions in terms of the Schur polynomials.

The Drinfeld-Sokolov hierarchies are characterized by the Heisenberg subalgebras, that is maximal nilpotent subalgebras, of the affine Lie algebras.
And the isomorphism classes of the Heisenberg subalgebras are in one-to-one correspondence with the conjugacy classes of the finite Weyl group {\rm\cite{KP}}.
In this article, we choose the {\it regular} conjugacy classes of $W(A_n)$ and consider their associated hierarchies, called {\it type I hierarchies} \cite{GHM}.
In the notation of \cite{DF}, the regular conjugacy classes of $W(A_n)$ correspond to the partitions $(p,\ldots,p)$ and $(p,\ldots,p,1)$.
For the derivation of fourth order Painlev\'{e} systems, we investigate the partitions $(2,2)$, $(3,1)$, $(4,1)$, $(2,2,1)$ and $(3,3)$; {\it see Table 2}.\DSPainleveResult

One of impotant results in this article is the derivation of a new Painlev\'{e} system.
It is expressed as a Hamiltonian system
\begin{equation}\label{Eq:CP6}
	\frac{dq_i}{dt} = \frac{\partial H_c}{\partial p_i},\quad
	\frac{dp_i}{dt} = -\frac{\partial H_c}{\partial q_i}\quad (i=1,2),
\end{equation}
with a Coupled Hamiltonian
\begin{equation}\begin{split}\label{Eq:CP6_Ham}
	t(t-1)H_c &= H_{\rm{VI}}(q_1,p_1;\alpha_2,\alpha_0+\alpha_4,
	\alpha_3+\alpha_5-\eta,\eta\alpha_1)\\
	&\quad + H_{\rm{VI}}(q_2,p_2;\alpha_0+\alpha_2,\alpha_4,
	\alpha_1+\alpha_3-\eta,\eta\alpha_5)\\
	&\quad
	+ (q_1-t)(q_2-1)\left\{(q_1p_1+\alpha_1)p_2+p_1(p_2q_2+\alpha_5)\right\}.
\end{split}\end{equation}
This system admits affine Weyl group symmetry of type $A_5^{(1)}$; see Appendix \ref{Sec:AffWey}.
On the other hand, the system $\mathcal{H}^{D_6^{(1)}}$ admits one of type $D_6^{(1)}$.
The relation between those two coupled Painlev\'{e} VI systems is not clarified.

\begin{rem}
For the partition $(1,\ldots,1)$ of $n+2$, we have the Garnier system in $n$-variables {\rm\cite{KK2}}.
Also for each partition $(5,1)$ and $(2,2,2)$, a system of sixth order is derived{\rm;} we do not give the explicit formula here.
Thus we conjecture that any more fourth order Painlev\'{e} system do not arise from the type I hierarchy.
\end{rem}

This article is organized as follows.
In Section \ref{Sec:AffLie}, we recall the affine Lie algebra of type $A^{(1)}_n$ and realize it in a flamework of a central extension of the loop algebra $\mf{sl}_{n+1}[z,z^{-1}]$.
In Section \ref{Sec:Heisenberg}, the Heisenberg subalgebra of $\wh{\mf{sl}}_{n+1}$ corresponding to the partition $\mathbf{n}$ is introduced.
In Section \ref{Sec:D-S}, we formulate the Drinfeld-Sokolov hierarchies and their similarity reductions.
In Section \ref{Sec:Deri_CP6} and \ref{Sec:Deri_Others}, the Painlev\'{e} systems are derived from the Drinfeld-Sokolov hierarchies.
In Appendix \ref{Sec:Lax}, we give explicit descriptions of Lax pairs by means of a bases of $\wh{\mathfrak{sl}}_{n+1}$.
In Appendix \ref{Sec:AffWey}, we discuss a group of symmetries for the system \eqref{Eq:CP6} with \eqref{Eq:CP6_Ham}.

\section{Affine Lie algebra}\label{Sec:AffLie}

In this section, we recall the affine Lie algebra of type $A^{(1)}_n$ and realize it in a flamework of a central extension of the loop algebra $\mf{sl}_{n+1}[z,z^{-1}]$.

In the notation of \cite{Kac}, the affine Lie algebra $\mf{g}=\mf{g}(A^{(1)}_n)$ is generated by the Chevalley generators $e_i,f_i,\alpha_i^{\vee}$ $(i=0,\ldots,n)$ and the scaling element $d$ with the fundamental relations
\[\begin{split}
	&(\mathrm{ad}e_i)^{1-a_{i,j}}(e_j)=0,\quad
	(\mathrm{ad}f_i)^{1-a_{i,j}}(f_j)=0\quad (i\neq j),\\
	&[\alpha_i^{\vee},\alpha_j^{\vee}]=0,\quad
	[\alpha_i^{\vee},e_j]=a_{i,j}e_j,\quad
	[\alpha_i^{\vee},f_j]=-a_{i,j}f_j,\quad
	[e_i,f_j]=\delta_{i,j}\alpha_i^{\vee},\\
	&[d,\alpha_i^{\vee}]=0,\quad [d,e_i]=\delta_{i,0}e_0,\quad
	[d,f_i]=-\delta_{i,0}f_0,
\end{split}\]
for $i,j=0,\ldots,n$.
The generalized Cartan matrix $A=\left[a_{i,j}\right]_{i,j=0}^{n}$ for $\mf{g}$ is defined by
\[\begin{array}{llll}
	a_{i,i}=2& (i=0,\ldots,n),\\[4pt]
	a_{i,i+1}=a_{n,0}=a_{i+1,i}=a_{0,n}=-1& (i=0,\ldots,n-1),\\[4pt]
	a_{i,j}=0& (\text{otherwise}).
\end{array}\]
We denote the Cartan subalgebra of $\mf{g}$ by
\[
	\mf{h} = \mathbb{C}\alpha_0^{\vee}\oplus\mathbb{C}\alpha_1^{\vee}
	\oplus\cdots\oplus\mathbb{C}\alpha_n^{\vee}\oplus\mathbb{C}d
	= \mf{h}'\oplus\mathbb{C}d.
\]
The normalized invariant form $(\cdot|\cdot):\mf{g}\times\mf{g}\to\mathbb{C}$ is determined by the conditions
\[\begin{array}{lll}
	(\alpha_i^{\vee}|\alpha_j^{\vee}) = a_{i,j},& (e_i|f_j) = \delta_{i,j},&
	(\alpha_i^{\vee}|e_j) = (\alpha_i^{\vee}|f_j) = 0,\\[4pt]
	(d|d) = 0,& (d|\alpha_j^{\vee}) = \delta_{0,j},& (d|e_j) = (d|f_j) = 0,
\end{array}\]
for $i,j=0,\ldots,n$.

Let $\mf{n}_{+}$ and $\mf{n}_{-}$ be the subalgebras of $\mf{g}$ generated by $e_i$ and $f_i$ $(i=0,\ldots,n)$ respectively.
Then the Borel subalgebra $\mf{b}_{+}$ of $\mf{g}$ is defined by $\mf{b}_{+}=\mf{h}\oplus\mf{n}_{+}$.
Note that we have the triangular decomposition
\[
	\mf{g} = \mf{n}_{-}\oplus\mf{h}\oplus\mf{n}_{+}
	= \mf{n}_{-}\oplus\mf{b}_{+}.
\]
The corresponding infinite demensional groups are defined by
\[
	N_{\pm} = \exp(\mf{n}_{\pm}^*),\quad H = \exp(\mf{h}'),\quad
	B_{+} = HN_{+},
\]
where $\mf{n}_{\pm}^*$ are completions of $\mf{n}_{\pm}$ respectively.

Let $\mathbf{s}=(s_0,\ldots,s_n)$ be a vector of non-negative integers.
We consider a gradation $\mf{g}=\bigoplus_{k\in\mathbb{Z}}\mf{g}_k(\mathbf{s})$ of type $\mathbf{s}$ by setting
\[
	\deg\mf{h}=0,\quad \deg e_i=s_i,\quad \deg f_i=-s_i\quad
	(i=0,\ldots,n).
\]
With an element $\vartheta(\mathbf{s})\in\mf{h}$ such that
\[
	(\vartheta(\mathbf{s})|\alpha_i^{\vee}) = s_i\quad (i=0,\ldots,n),
\]
this gradation is defined by
\[
	\mf{g}_k(\mathbf{s}) = \left\{x\in\mf{g}\bigm|
	[\vartheta(\mathbf{s}),x]=kx\right\}\quad (k\in\mathbb{Z}).
\]
We denote by
\[
	\mf{g}_{<k}(\mathbf{s}) = \bigoplus_{l<k}\mf{g}_l(\mathbf{s}),\quad
	\mf{g}_{\geq k}(\mathbf{s}) = \bigoplus_{l\geq k}\mf{g}_l(\mathbf{s}).
\]
Note that a gradation $\mathbf{s}_p=(1,\ldots,1)$, called {\it the principal gradation}, implies
\[
	\mf{g}_{<0}(\mathbf{s}_p) = \mf{n}_{-},\quad
	\mf{g}_{\geq0}(\mathbf{s}_p) = \mf{b}_{+}.
\]

The affine Lie algebra $\mf{g}$ can be identified with
\[
	\wh{\mf{sl}}_{n+1} = \mf{sl}_{n+1}[z,z^{-1}]\oplus\mathbb{C}z\frac{d}{dz}
	\oplus\mathbb{C}K,
\]
where $K$ is a canonical central element.
In a flamework of $\wh{\mf{sl}}_{n+1}$, the Chevalley generators and the scaling element are given by
\[\begin{split}
	&e_i = E_{i,i+1},\quad f_i = E_{i+1,i},\quad
	\alpha^{\vee}_i = E_{i,i} - E_{i+1,i+1}\quad (i=1,\ldots,n),\\
	&e_0 = zE_{n+1,1},\quad f_0 = z^{-1}E_{1,n+1},\quad
	\alpha^{\vee}_0 = E_{n+1,n+1} - E_{1,1} + K,\quad d = z\frac{d}{dz},
\end{split}\]
where $E_{i,j}=\left(\delta_{i,r}\delta_{j,s}\right)_{r,s=1}^{n+1}$ are matrix units.
The Lie bracket is defined by
\[\begin{split}
	&[z^kX,z^lY] = z^{k+l}(XY-YX) + k\delta_{k+l,0}\mathrm{tr}(XY)K,
\end{split}\]
where $X,Y\in\mf{sl}_{n+1}$.

\section{Heisenberg subalgebra}\label{Sec:Heisenberg}

For type $A_n^{(1)}$, the isomorphism classes of the Heisenberg subalgebras are in one-to-one correspondence with the partitions of $n+1$.
In this section, we introduce the Heisenberg subalgebra of $\wh{\mf{sl}}_{n+1}$ corresponding to the partition $\mathbf{n}$ following the manner in \cite{KL}.

Let $\mathbf{n}=(n_1,n_2,\ldots,n_r,n_{r+1},\ldots,n_s)$ be a partition of $n+1$ with $n_1\geq n_2\geq\ldots\geq n_r>n_{r+1}=\ldots=n_s=1$.
Consider a partition of matrix corresponding to $\mathbf{n}$
\[
	\begin{bmatrix}
		B_{11}& B_{12}& \cdots& B_{1s}\\
		B_{21}& B_{22}& \cdots& B_{2s}\\
		\vdots& \vdots& \ddots& \vdots\\
		B_{s1}& B_{s2}& \cdots& B_{ss}
	\end{bmatrix},
\]
where each block $B_{ij}$ is an $n_i\times n_j$-matrix.
With this blockform, we define matricies $\Lambda_i'\in\wh{\mf{sl}}_{n+1}$ $(i=1,\ldots,r)$ by
\[
	\Lambda_i' = \begin{bmatrix}
		O& & \cdots& & O\\ & & & & \\ \vdots& & B_{ii}& & \vdots\\
		& & & & \\ O& & \cdots& & O
	\end{bmatrix},\quad
	B_{ii} = \begin{bmatrix}
		0& 1& 0& \cdots& 0\\ 0& 0& 1& & 0\\ \vdots& \vdots& & \ddots& \\
		0& 0& 0& & 1\\ z& 0& 0& \cdots& 0
	\end{bmatrix},
\]
diagonal matricies $H_j'\in\wh{\mf{sl}}_{n+1}$ $(i=j,\ldots,s-1)$ by
\[
	H_j' = n_{j+1}z^{-1}(\Lambda_j')^{n_j}
	- n_jz^{-1}(\Lambda_{j+1}')^{n_{j+1}},
\]
and a diagonal matrix $\eta_{\mathbf{n}}'\in\wh{\mf{sl}}_{n+1}$ by
\[
	B_{ii} = \frac{1}{2n_i}\mathrm{diag}(n_i-1,n_i-3,\ldots,-n_i+1)\quad
	(i=1,\ldots,r).
\]

Denoting the matrix $\eta_{\mathbf{n}}'$ by $\mathrm{diag}(\eta_1',\eta_2',\ldots,\eta_{n+1}')$, we consider a permutation
\[
	\sigma = \left(\begin{array}{llll}
		\eta_1'& \eta_2'& \ldots& \eta_{n+1}'\\[4pt]
		\eta_1& \eta_2& \ldots& \eta_{n+1}
	\end{array}\right),
\]
such that $\eta_1\geq\eta_2\geq\ldots\geq\eta_{n+1}$.
This permutation can be lifted to the transformation $\sigma$ acting on the matricies $\Lambda_i'$ and $H_j'$.
We set
\[
	\Lambda_i = \sigma(\Lambda_i')\quad (i=1,\ldots,r),\quad
	H_j = \sigma(H_j')\quad (j=1,\ldots,s-1).
\]
Then the Heisenberg subalgebra of $\widehat{\mf{sl}}_{n+1}$ corresponding to the partition $\mathbf{n}$ is defined by
\[
	\mf{s}_\mathbf{n} = \bigoplus_{i=1}^{r}
	\bigoplus_{k\in\mathbb{Z}\setminus n_i\mathbb{Z}}\mathbb{C}\Lambda_i^k
	\oplus\bigoplus_{j=1}^{s-1}\bigoplus_{k\in\mathbb{Z}\setminus\{0\}}
	\mathbb{C}z^kH_j\oplus\mathbb{C}K.
\]

Let $N_{\mathbf{n}}'$ be the least common multiple of $n_1,\ldots,n_s$.
Also let
\[
	N_{\mathbf{n}} = \left\{\begin{array}{ll}
		N_{\mathbf{n}}'&
		\text{if $\ds N_{\mathbf{n}}'\left(\frac{1}{n_i}+\frac{1}{n_j}\right)
		\in2\mathbb{Z}$ for $\forall(i,j)$}\\[12pt]
		2N_{\mathbf{n}}'& \text{otherwise}
	\end{array}\right..
\]
We consider a operator corresponding to $\mathbf{n}$
\[
	\vartheta_{\mathbf{n}}
	= N_{\mathbf{n}}\left(z\frac{d}{dz}+\mathrm{ad}\eta_{\mathbf{n}}\right),
\]
where $\eta_{\mathbf{n}}=\sigma(\eta_{\mathbf{n}}')$.
Then the operator $\vartheta_{\mathbf{n}}$ implies a gradation $\mathbf{s}=(s_0,\ldots,s_n)$ as follows:
\[
	\vartheta_{\mathbf{n}}(e_i) = s_ie_i\quad (i=0,\ldots,n).
\]
Note that the Heisenberg subalgebra $\mf{s}_\mathbf{n}$ admits the gradation $\mathbf{s}$ defined by $\vartheta_{\mathbf{n}}$.

\section{Drinfeld-Sokolov hierarchy}\label{Sec:D-S}

In this section, we formulate the Drinfeld-Sokolov hierarchy associated with the Heisenberg subalgebra $\mf{s}_\mathbf{n}$.
Its similarity reduction is also formulated.

Let $\Lambda_i$ and $H_j$ be the generators for $\mf{s}_\mathbf{n}$ given in Section \ref{Sec:Heisenberg}.
Introducing time variables $t_{i,k}$ $(i=1,\ldots,r;k\in\mathbb{N})$, we consider an $N_{-}B_{+}$-valued function $G=G(t_{1,1},t_{1,2},\ldots)$ defined by
\[
	G = \exp\left(\sum_{i=1}^r\sum_{k=1}^{\infty}t_{i,k}\Lambda_i^k\right)G(0).
\]
Here we assume the $\mathbf{n}$-reduced condition
\[
	t_{i,l}=0\quad (i=1,\ldots,r;l\in n_i\mathbb{N}).
\]
Then we have a system of partial differential equations
\begin{equation}\label{Eq:DS_exp}
	\partial_{i,k}(G) = \Lambda_i^kG\quad (i=1,\ldots,r;k\in\mathbb{N}),
\end{equation}
where $\partial_{i,k}=\partial/\partial t_{i,k}$
Via the trianglar decomposition
\[
	G = W^{-1}Z,\quad W\in N_{-},\quad Z\in B_{+},
\]
the system \eqref{Eq:DS_exp} implies {\it a Sato equation}
\begin{equation}\label{Eq:Sato}
	\partial_{i,k}(W) = B_{i,k}W - W\Lambda_i^k\quad
	(i=1,\ldots,r;k\in\mathbb{N}),
\end{equation}
where $B_{i,k}$ stands for the $b_{+}$-component of $W\Lambda_i^kW^{-1}$.
The compatibility condition of \eqref{Eq:Sato} gives the Drinfeld-Sokolov hierarchy
\begin{equation}\label{Eq:DS}
	\left[\partial_{i,k}-B_{i,k},\partial_{j,l}-B_{j,l}\right] = 0\quad
	(i,j=1,\ldots,r;k,l\in\mathbb{N}).
\end{equation}

Under the system \eqref{Eq:Sato}, we consider an equation
\begin{equation}\label{Eq:Sato_SR}
	(\vartheta_{\mathbf{n}}-\mathrm{ad}\rho)(W)
	= \sum_{i=1}^r\sum_{k=1}^{\infty}d_ikt_{i,k}\partial_{i,k}(W),
\end{equation}
where $d_i=\deg\Lambda_i$ $(i=1,\ldots,r)$ and $\rho=\sum_{j=1}^{s-1}\rho_jH_j$.
Note that each $\rho_j$ is independent of time vatiables $t_{i,k}$.
The compatibility condition of \eqref{Eq:Sato} and \eqref{Eq:Sato_SR} gives
\begin{equation}\label{Eq:DS_SR}
	\left[\vartheta_{\mathbf{n}}-M,\partial_{i,k}-B_{i,k}\right] = 0\quad
	(i=1,\ldots,r;k\in\mathbb{N}),
\end{equation}
where
\[
	M = \rho + \sum_{i=1}^r\sum_{k=1}^{\infty}d_ikt_{i,k}B_{i,k}.
\]
We call the systems \eqref{Eq:DS} and \eqref{Eq:DS_SR} a similarity reduction of the Drinfeld-Sokolov hierarchy.

\begin{rem}\label{Rem:Lax}
The similarity reduction can be regarded as the compatibility condition of a Lax form
\[
	\partial_{i,k}(\Psi) = B_{i,k}\Psi\quad (i=1,\ldots,r;k\in\mathbb{N}),\quad
	\vartheta_{\mathbf{n}}(\Psi) = M\Psi.
\]
Here an $N_{-}B_{+}$-valued function $\Psi$ is given by
\[
	\Psi = W\exp\left(\sum_{i=1}^r\sum_{k=1}^{\infty}t_{i,k}\Lambda_i^k\right).
\]
\end{rem}

\section{Derivation of Coupled $P_{\rm{VI}}$}\label{Sec:Deri_CP6}

In this section, we derive the Painlev\'{e} system \eqref{Eq:CP6} with \eqref{Eq:CP6_Ham} from the Drinfeld-Sokolov hierarchies for $\mf{s}_{(3,3)}$ and $\mf{s}_{(2,2,1)}$ by similarity reductions.

\subsection{For the partition $(3,3)$}\label{Sec:System33}

At first, we define the Heisenberg subalgebra $\mf{s}_{(3,3)}$ of $\mf{g}(A^{(1)}_5)$.
Let
\[
	\Lambda_1 = e_{1,2} + e_{3,4} + e_{5,0},\quad
	\Lambda_2 = e_{0,1} + e_{2,3} + e_{4,5},\quad
	H_1 = \alpha^{\vee}_1 + \alpha^{\vee}_3 + \alpha^{\vee}_5,
\]
where
\[
	e_{i_1,i_2,\ldots,i_{n-1},i_n} = \mathrm{ad}e_{i_1}\mathrm{ad}e_{i_2}
	\ldots\mathrm{ad}e_{i_{n-1}}(e_{i_n}).
\]
Then we have
\[
	\mf{s}_{(3,3)} = \bigoplus_{k\in\mathbb{Z}\setminus3\mathbb{Z}}
	\mathbb{C}\Lambda_1^k\oplus\bigoplus_{k\in\mathbb{Z}\setminus3\mathbb{Z}}
	\mathbb{C}\Lambda_2^k\oplus\bigoplus_{k\in\mathbb{Z}\setminus\{0\}}
	\mathbb{C}z^kH_1\oplus\mathbb{C}K.
\]
The grade operator for $\mf{s}_{(3,3)}$ is given by
\[
	\vartheta_{(3,3)} = 3\left(z\frac{d}{dz}+\mathrm{ad}\eta_{(3,3)}\right),
\]
where
\[
	\eta_{(3,3)} = \frac{1}{3}(\alpha^{\vee}_1+2\alpha^{\vee}_2+2\alpha^{\vee}_3
	+2\alpha^{\vee}_4+\alpha^{\vee}_5).
\]
It follows that $\mf{s}_{(3,3)}$ admits the gradation of type $\mathbf{s}=(1,0,1,0,1,0)$, namely
\[
	\vartheta_{(3,3)}(e_i) = e_i\quad (i=0,2,4),\quad
	\vartheta_{(3,3)}(e_j) = 0\quad (j=1,3,5).
\]
Note that
\[
	\mf{g}_{\geq0}(1,0,1,0,1,0) = \mathbb{C}f_1\oplus\mathbb{C}f_3
	\oplus\mathbb{C}f_5\oplus\mf{b}_{+}.
\]

We now assume $t_{2,1}=1$ and $t_{1,k}=t_{2,k}=0$ $(k\geq2)$.
Then the similarity reduction \eqref{Eq:DS} and \eqref{Eq:DS_SR} for $\mf{s}_{(3,3)}$ is expressed as
\begin{equation}\label{Eq:DS_SR_33_b}
	\left[\vartheta_{(3,3)}-M,\partial_{1,1}-B_{1,1}\right] = 0.
\end{equation}
Here the $\mf{b}_{+}$-valued functions $M$ and $B_{1,1}$ are defined by
\begin{equation}\begin{split}\label{Eq:DS_SR_33_b_BM}
	M &= \vartheta_{(3,3)}(W)W^{-1}
	+ W(\rho_1H_1+t_{1,1}\Lambda_1+\Lambda_2)W^{-1},\\
	B_{1,1} &= \partial_{1,1}(W)W^{-1} + W\Lambda_1W^{-1},
\end{split}\end{equation}
where $W$ is an $N_{-}$-valued function; its explicit formula is given below.
In the following, we derive the Painlev\'{e} system from the system \eqref{Eq:DS_SR_33_b} with \eqref{Eq:DS_SR_33_b_BM}.

We denote by
\[
	W = \exp(\omega_0)\exp(\omega_{-1})\exp(\omega_{<-1}),
\]
where
\[\begin{split}
	\omega_0 &= -w_1f_1 - w_3f_3 - w_5f_5,\\
	\omega_{-1} &= -w_0f_0 - w_2f_2 - w_4f_4 - w_{0,1}f_{0,1} - w_{1,2}f_{1,2}
	- w_{2,3}f_{2,3} - w_{3,4}f_{3,4}\\
	&\quad - w_{4,5}f_{4,5} - w_{5,0}f_{5,0} - w_{1,2,3}f_{1,2,3}
	- w_{3,4,5}f_{3,4,5} - w_{5,0,1}f_{5,0,1},
\end{split}\]
and $\omega_{<-1}\in\mf{g}_{<-1}(1,0,1,0,1,0)$.
Then the $\mf{b}_{+}$-valued function $M$ is described as
\[\begin{split}
	M &= \kappa_0\alpha^{\vee}_0 + \kappa_1\alpha^{\vee}_1
	+ \kappa_2\alpha^{\vee}_2 + \kappa_3\alpha^{\vee}_3
	+ \kappa_4\alpha^{\vee}_4 + \kappa_5\alpha^{\vee}_5 - (t_{1,1}w_5-w_1)e_0
	+ \varphi_1e_1\\
	&\quad - (t_{1,1}w_1-w_3)e_2 + \varphi_3e_3 - (t_{1,1}w_3-w_5)e_4
	+ \varphi_5e_5 + t_{1,1}\Lambda_1 + \Lambda_2,\\
\end{split}\]
with dependent variables
\[
	\varphi_1 = t_{1,1}w_2 - w_0,\quad \varphi_3= t_{1,1}w_4 - w_2,\quad
	\varphi_5 = t_{1,1}w_0 - w_4,
\]
and parameters
\[\begin{split}
	&\kappa_0 = -t_{1,1}w_{5,0} - w_{0,1},\quad
	\kappa_1 = t_{1,1}(w_1w_2-w_{1,2}) - (w_0w_1+w_{0,1}) + \rho_1,\\
	&\kappa_2 = -t_{1,1}w_{1,2} - w_{2,3},\quad
	\kappa_3 = t_{1,1}(w_3w_4-w_{3,4}) - (w_2w_3+w_{2,3}) + \rho_1,\\
	&\kappa_4 = -t_{1,1}w_{3,4} - w_{4,5},\quad
	\kappa_5 = t_{1,1}(w_0w_5-w_{5,0}) - (w_4w_5+w_{4,5}) + \rho_1.
\end{split}\]
Note that
\[
	\partial_{1,1}(\kappa_i) = 0\quad (i=0,\ldots,5).
\]
We also remark that
\[
	w_1\varphi_1 + w_3\varphi_3 + w_5\varphi_5 + \kappa_0 - \kappa_1
	+ \kappa_2 - \kappa_3 + \kappa_4 - \kappa_5 + 3\rho_1 = 0.
\]
The $\mf{b}_{+}$-valued function $B_{1,1}$ is described as
\[\begin{split}
	B_{1,1} &= u_0K + (u_1+w_1x_1)\alpha^{\vee}_1 + u_2\alpha^{\vee}_2
	+ (u_3+w_3x_3)\alpha^{\vee}_3 + u_4\alpha^{\vee}_4\\
	&\quad + w_5x_5\alpha^{\vee}_5 - w_5e_0 + x_1e_1 - w_1e_2 + x_3e_3
	- w_3e_4 + x_5e_5 + \Lambda_1,
\end{split}\]
where
\[\begin{split}
	&u_1 = \frac{-2w_1\varphi_1+w_3\varphi_3+w_5\varphi_5
	-2\kappa_0+2\kappa_1+\kappa_2-\kappa_3+\kappa_4-\kappa_5}{3t_{1,1}},\\
	&u_2 = -\frac{w_1\varphi_1+\kappa_0-\kappa_1+\rho_1}{t_{1,1}},\\
	&u_3 = \frac{-w_1\varphi_1-w_3\varphi_3+2w_5\varphi_5
	-\kappa_0+\kappa_1-\kappa_2+\kappa_3+2\kappa_4-2\kappa_5}{3t_{1,1}},\\
	&u_4 = \frac{w_5\varphi_5+\kappa_4-\kappa_5+\rho_1}{t_{1,1}},\quad
	x_1 = \frac{t_{1,1}^2\varphi_1+t_{1,1}\varphi_5+\varphi_3}{t_{1,1}^3-1},\\
	&x_3 = \frac{t_{1,1}^2\varphi_3+t_{1,1}\varphi_1+\varphi_5}
	{t_{1,1}^3-1},\quad
	x_5 = \frac{t_{1,1}^2\varphi_5+t_{1,1}\varphi_3+\varphi_1}{t_{1,1}^3-1}.
\end{split}\]
Hence the system \eqref{Eq:DS_SR_33_b} with \eqref{Eq:DS_SR_33_b_BM} can be expressed as a system of ordinary differential equations in terms of the variabes $\varphi_1,\varphi_5,w_1,w_3,w_5$; we do not give its explicit formula.

Let
\[
	q_1 = \frac{w_1}{t_{1,1}^2w_3},\quad
	p_1 = \frac{t_{1,1}^2w_3\varphi_1}{3},\quad
	q_2 = \frac{w_5}{t_{1,1}w_3},\quad
	p_2 = \frac{t_{1,1}w_3\varphi_5}{3},\quad t = \frac{1}{t_{1,1}^3}.
\]
We also set
\[\begin{split}
	&\alpha_0 = \frac{1}{3}(1-2\kappa_0+\kappa_1+\kappa_5),\quad
	\alpha_1 = \frac{1}{3}(\kappa_0-2\kappa_1+\kappa_2),\\
	&\alpha_2 = \frac{1}{3}(1+\kappa_1-2\kappa_2+\kappa_3),\quad
	\alpha_3 = \frac{1}{3}(\kappa_2-2\kappa_3+\kappa_4),\\
	&\alpha_4 = \frac{1}{3}(1+\kappa_3-2\kappa_4+\kappa_5),\quad
	\alpha_5 = \frac{1}{3}(\kappa_0+\kappa_4-2\kappa_5),
\end{split}\]
and
\[
	\eta = \rho_1 + \frac{1}{2}(\alpha_1+\alpha_3+\alpha_5).
\]
Then we have

\begin{thm}
The system \eqref{Eq:DS_SR_33_b} with \eqref{Eq:DS_SR_33_b_BM} gives the Painlev\'{e} system \eqref{Eq:CP6} with \eqref{Eq:CP6_Ham}.
Furthermore, $w_3$ satisfies the completely integrable Pfaffian equation
\[\begin{split}
	t(t-1)\frac{d}{dt}\log w_3 &= -(q_1-1)(q_1-t)p_1 - (q_2-1)(q_2-t)p_2\\
	&\quad - \alpha_1q_1 - \alpha_5q_2
	+ \frac{1}{3}(\alpha_1+\alpha_2-\alpha_3-\alpha_4+2\eta)t\\
	&\quad - \frac{1}{3}(\alpha_1+\alpha_2+2\alpha_3-\alpha_4-4\eta).
\end{split}\]
\end{thm}

\subsection{For the partition $(2,2,1)$}\label{Sec:System221}

The Heisenberg subalgebra $\mf{s}_{(2,2,1)}$ of $\mf{g}(A^{(1)}_4)$ is defined by
\[
	\mf{s}_{(2,2,1)} = \bigoplus_{k\in\mathbb{Z}\setminus2\mathbb{Z}}
	\mathbb{C}\Lambda_1^k\oplus\bigoplus_{k\in\mathbb{Z}\setminus2\mathbb{Z}}
	\mathbb{C}\Lambda_2^k\oplus\bigoplus_{k\in\mathbb{Z}\setminus\{0\}}
	\mathbb{C}z^kH_1\oplus\bigoplus_{k\in\mathbb{Z}\setminus\{0\}}
	\mathbb{C}z^kH_2\oplus\mathbb{C}K,
\]
with
\[\begin{array}{ll}
	\Lambda_1 = e_{4,0} + e_{1,2,3},& \Lambda_2 = e_{0,1} + e_{2,3,4},\\[4pt]
	H_1 = \alpha^{\vee}_1 + \alpha^{\vee}_2 - \alpha^{\vee}_3,&
	H_2 = -\alpha^{\vee}_2 + \alpha^{\vee}_3 + \alpha^{\vee}_4.
\end{array}\]
The subalgebra $\mf{s}_{(2,2,1)}$ admits the gradation of type $\mathbf{s}=(2,0,1,1,0)$ with the grade operator
\[
	\vartheta_{(2,2,1)}
	= 4\left(z\frac{d}{dz}+\mathrm{ad}\eta_{(2,2,1)}\right),\quad
	\eta_{(2,2,1)} = \frac{1}{4}
	(\alpha^{\vee}_1+2\alpha^{\vee}_2+2\alpha^{\vee}_3+\alpha^{\vee}_4).
\]
Note that
\[
	\mf{g}_{\geq0}(2,0,1,1,0)
	= \mathbb{C}f_1\oplus\mathbb{C}f_4\oplus\mf{b}_{+}.
\]

We now assume $t_{1,2}=1$ and $t_{1,k}=t_{2,k}=0$ $(k\geq3)$.
Then the similarity reduction \eqref{Eq:DS_SR} for $\mf{s}_{(2,2,1)}$ is expressed as
\begin{equation}\label{Eq:DS_SR_221_b}
	\left[\vartheta_{(2,2,1)}-M,\partial_{1,1}-B_{1,1}\right] = 0,
\end{equation}
with
\begin{equation}\begin{split}\label{Eq:DS_SR_221_b_BM}
	M &= \vartheta_{(2,2,1)}(W)W^{-1}
	+ W(\rho_1H_1+\rho_2H_2+2t_{1,1}\Lambda_1+2\Lambda_2)W^{-1},\\
	B_{1,1} &= \partial_{1,1}(W)W^{-1} + W\Lambda_1W^{-1}.
\end{split}\end{equation}

Let
\[
	W = \exp(\omega_0)\exp(\omega_{-1})\exp(\omega_{-2})\exp(\omega_{<-2}),
\]
where
\[\begin{split}
	\omega_0 &= -w_1f_1 - w_4f_4,\\
	\omega_{-1} &= -w_2f_2 - w_3f_3 - w_{1,2}f_{1,2} - w_{3,4}f_{3,4},\\
	\omega_{-2} &= -w_0f_0 - w_{0,1}f_{0,1} - w_{2,3}f_{2,3} - w_{4,0}f_{4,0}\\
	&\quad - w_{1,2,3}f_{1,2,3} - w_{2,3,4}f_{2,3,4} - w_{4,0,1}f_{4,0,1}
	- w_{1,2,3,4}f_{1,2,3,4},
\end{split}\]
and $\omega_{<-2}\in\mf{g}_{<-2}(2,0,1,1,0)$.
Then the system \eqref{Eq:DS_SR_221_b_BM} gives explicit formulas of $M,B_{1,1}$ as follows:
\[\begin{split}
	M &= \kappa_0\alpha^{\vee}_0 + \kappa_1\alpha^{\vee}_1
	+ \kappa_2\alpha^{\vee}_2 + \kappa_3\alpha^{\vee}_3
	+ \kappa_4\alpha^{\vee}_4 + 2(w_1-t_{1,1}w_4)e_0\\
	&\quad + \varphi_1e_1 + (\varphi_2-w_1\varphi_{1,2})e_2
	+ (\varphi_3+w_4\varphi_{3,4})e_3 + \varphi_4e_4\\
	&\quad + \varphi_{1,2}e_{1,2} + 2(t_{1,1}w_1-w_4)e_{2,3}
	- \varphi_{3,4}e_{3,4} + 2t_{1,1}\Lambda_1 + 2\Lambda_2,\\
	B_{1,1} &= u_0K + (u_2+w_1x_1)\alpha^{\vee}_1 + u_2\alpha^{\vee}_2
	+ u_3\alpha^{\vee}_3 + w_4x_4\alpha^{\vee}_4 - w_4e_0\\
	&\quad + x_1e_1 - w_1x_{1,2}e_2 + \frac{\varphi_3}{2t_{1,1}}e_3 + x_4e_4
	+ x_{1,2}e_{1,2} - w_1e_{2,3} + \Lambda_1,
\end{split}\]
where
\[\begin{split}
	&\varphi_1 = -2w_0 + t_{1,1}w_2w_3 - 2t_{1,1}w_{2,3},\quad
	\varphi_2 = -2w_{3,4},\quad \varphi_3 = 2t_{1,1}w_{1,2},\\
	&\varphi_4 = 2t_{1,1}w_0 + w_2w_3 + 2w_{2,3},\quad
	\varphi_{1,2} = 2t_{1,1}w_3,\quad \varphi_{3,4} = -2w_2,
\end{split}\]
and
\[\begin{split}
	&u_2 = -\frac{w_1\varphi_1+\kappa_0-\kappa_1+\rho_1}{2t_{1,1}},\quad
	u_3 = \frac{w_4\varphi_4+\kappa_3-\kappa_4+\rho_1}{2t_{1,1}},\\
	&x_1 = \frac{(t_{1,1}\varphi_1+\varphi_4)\varphi_3
	+(w_1\varphi_1+w_4\varphi_4+\kappa_0-\kappa_1+\kappa_3-\kappa_4+2\rho_1)
	\varphi_{3,4}}{2(t_{1,1}^2-1)\varphi_3},\\
	&x_4 = \frac{(\varphi_1+t_{1,1}\varphi_4)\varphi_3+t_{1,1}
	(w_1\varphi_1+w_4\varphi_4+\kappa_0-\kappa_1+\kappa_3-\kappa_4+2\rho_1)
	\varphi_{3,4}}{2(t_{1,1}^2-1)\varphi_3},\\
	&x_{1,2} = \frac{w_1\varphi_1+w_4\varphi_4+\kappa_0-\kappa_1+\kappa_3
	-\kappa_4+2\rho_1}{\varphi_3}.
\end{split}\]
Note that $\kappa_0,\ldots,\kappa_4$ are constants.
We also remark that
\[\begin{split}
	&\varphi_2\varphi_{3,4} + 2(w_1\varphi_1+w_4\varphi_4+\kappa_0-\kappa_1
	+\kappa_2-\kappa_4+2\rho_2) = 0,\\
	&\varphi_3\varphi_{1,2} - 2t_{1,1}(w_1\varphi_1+w_4\varphi_4+\kappa_0
	-\kappa_1+\kappa_3-\kappa_4+2\rho_1) = 0.
\end{split}\]
Hence the system \eqref{Eq:DS_SR_221_b} can be expressed as a system of ordinary differential equations in terms of the variables $\varphi_1,\varphi_3,\varphi_4,\varphi_{3,4},w_1,w_4$.

Let
\[\begin{split}
	&q_1 = -\frac{t_{1,1}^2\varphi_{3,4}w_4}{\varphi_3},\quad
	p_1 = -\frac{\varphi_3\varphi_4}{4t_{1,1}^2\varphi_{3,4}},\\
	&q_2 = -\frac{t_{1,1}\varphi_{3,4}w_1}{\varphi_3},\quad
	p_2 = -\frac{\varphi_3\varphi_1}{4t_{1,1}\varphi_{3,4}},\quad
	t = t_{1,1}^2.
\end{split}\]
We also set
\[\begin{split}
	&\alpha_0 = \frac{1}{4}(2-2\kappa_0+\kappa_1+\kappa_4),\quad
	\alpha_1 = \frac{1}{4}(\kappa_0+\kappa_3-2\kappa_4),\\
	&\alpha_2 = \frac{1}{4}(1+\kappa_2-2\kappa_3+\kappa_4),\quad
	\alpha_3 = \frac{1}{4}(-\kappa_2+\kappa_3+2\rho_1-2\rho_2),\\
	&\alpha_4 = \frac{1}{4}(1+\kappa_1-\kappa_2-2\rho_1+2\rho_2),\quad
	\alpha_5 = \frac{1}{4}(\kappa_0-2\kappa_1+\kappa_2),\\
	&\eta
	= \frac{1}{4}(2\kappa_0-2\kappa_1+2\kappa_3-2\kappa_4+3\rho_1-\rho_2).
\end{split}\]
Then we have

\begin{thm}
The system \eqref{Eq:DS_SR_221_b} with \eqref{Eq:DS_SR_221_b_BM} gives the Painlev\'{e} system \eqref{Eq:CP6} with \eqref{Eq:CP6_Ham}.
Furthermore, $\varphi_3$ and $\varphi_{3,4}$ satisfy the completely integrable Pfaffian equations
\[\begin{split}
	t(t-1)\frac{d}{dt}\log\varphi_3 &= -q_1(q_1-t)p_1 - q_2(q_2-t)p_2
	- \alpha_1q_1 - \alpha_5q_2\\
	&\quad + \frac{1}{4}
	(1+2\alpha_2-2\alpha_3-2\alpha_4-2\alpha_5+6\eta)t\\
	&\quad - \frac{1}{4}
	(1+2\alpha_2+2\alpha_3-2\alpha_4-2\alpha_5+2\eta),\\
	t(t-1)\frac{d}{dt}\log\varphi_{3,4} &= -(q_1-t)p_1 - (q_2-t)p_2 - \eta.
\end{split}\]
\end{thm}

\section{Derivation of other systems}\label{Sec:Deri_Others}

In this section, we discuss the derivation of the Painlev\'{e} systems for $\mf{s}_{(2,2)}$, $\mf{s}_{(3,1)}$ and $\mf{s}_{(4,1)}$ by a similar manner as in Section \ref{Sec:Deri_CP6}.

\subsection{For the partition $(2,2)$}\label{Sec:System22}

The Heisenberg subalgebra $\mf{s}_{(2,2)}$ of $\mf{g}(A^{(1)}_3)$ is defined by
\[
	\mf{s}_{(2,2)} = \bigoplus_{k\in\mathbb{Z}\setminus2\mathbb{Z}}
	\mathbb{C}\Lambda_1^k\oplus\bigoplus_{k\in\mathbb{Z}\setminus2\mathbb{Z}}
	\mathbb{C}\Lambda_2^k\oplus\bigoplus_{k\in\mathbb{Z}\setminus\{0\}}
	\mathbb{C}z^kH_1\oplus\mathbb{C}K,
\]
with
\[
	\Lambda_1 = e_{1,2} + e_{3,0},\quad
	\Lambda_2 = e_{0,1} + e_{2,3},\quad
	H_1 = \alpha^{\vee}_1 + \alpha^{\vee}_3.
\]
The subalgebra $\mf{s}_{(2,2)}$ admits the gradation of type $\mathbf{s}=(1,0,1,0)$ with the grade operator
\[
	\vartheta_{(2,2)}
	= 2\left(z\frac{d}{dz}+\mathrm{ad}\eta_{(2,2)}\right),\quad
	\eta_{(2,2)}
	= \frac{1}{2}(\alpha^{\vee}_1+2\alpha^{\vee}_2+\alpha^{\vee}_3).
\]
Note that
\[
	\mf{g}_{\geq0}(1,0,1,0) = \mathbb{C}f_1\oplus\mathbb{C}f_3\oplus\mf{b}_{+}.
\]

We now assume $t_{1,2}=1$ and $t_{1,k}=t_{2,k}=0$ $(k\geq3)$.
Then the similarity reduction \eqref{Eq:DS_SR} for $\mf{s}_{(2,2)}$ is expressed as
\begin{equation}\label{Eq:DS_SR_22_b}
	\left[\vartheta_{(2,2)}-M,\partial_{1,1}-B_{1,1}\right] = 0,
\end{equation}
with
\begin{equation}\begin{split}\label{Eq:DS_SR_22_b_BM}
	M &= \vartheta_{(2,2)}(W)W^{-1}
	+ W(\rho_1H_1+t_{1,1}\Lambda_1+\Lambda_2)W^{-1},\\
	B_{1,1} &= \partial_{1,1}(W)W^{-1} + W\Lambda_1W^{-1}.
\end{split}\end{equation}

Let
\[
	W = \exp(\omega_0)\exp(\omega_{-1})\exp(\omega_{<-1}),
\]
where
\[\begin{split}
	\omega_{0} &= -w_1f_1 - w_3f_3,\\
	\omega_{-1} &= -w_0f_0 - w_2f_2 - w_{0,2}f_{0,2} - w_{1,2}f_{1,2}\\
	&\quad - w_{2,3}f_{2,3} - w_{3,0}f_{3,0} - w_{1,2,3}f_{1,2,3}
	- w_{3,0,1}f_{3,0,1},
\end{split}\]
and $\omega_{<-1}\in\mf{g}_{<-1}(1,0,1,0)$.
Then the system \eqref{Eq:DS_SR_22_b_BM} gives explicit formulas of $M,B_{1,1}$ as follows:
\[\begin{split}
	M &= \kappa_0\alpha^{\vee}_0 + \kappa_1\alpha^{\vee}_1
	+ \kappa_2\alpha^{\vee}_2 + \kappa_3\alpha^{\vee}_3 + (w_1-t_{1,1}w_3)e_0\\
	&\quad + \varphi_1e_1 + (w_3-t_{1,1}w_1)e_2 + \varphi_3e_3
	+ t_{1,1}\Lambda_1 + \Lambda_2,\\
	B_{1,1} &= u_0K + u_1\alpha^{\vee}_1 + u_2\alpha^{\vee}_2
	+ w_3x_3\alpha^{\vee}_3 + w_1e_0 + x_1e_1 + w_3e_2 + x_3e_3 + \Lambda_1,
\end{split}\]
where
\[
	\varphi_1 = t_{1,1}w_2 - w_0,\quad \varphi_3 = t_{1,1}w_0 - w_2,
\]
and
\[\begin{split}
	u_1 &= \frac{w_1}{t_{1,1}}x_3
	- \frac{\kappa_0-\kappa_1+\rho_1}{t_{1,1}},\quad
	u_2 = \frac{w_3\varphi_3+\kappa_2-\kappa_3+\rho_1}{t_{1,1}},\\
	x_1 &= \frac{(w_1-t_{1,1}w_3)\varphi_3
	-(\kappa_0-\kappa_1+\kappa_2-\kappa_3+2\rho_1)t_{1,1}}{(t_{1,1}^2-1)w_1},\\
	x_3 &= \frac{(t_{1,1}w_1-w_3)\varphi_3
	-(\kappa_0-\kappa_1+\kappa_2-\kappa_3+2\rho_1)}{(t_{1,1}^2-1)w_1}.
\end{split}\]
Note that $\kappa_0,\ldots,\kappa_3$ are constants.
We also remark that
\[
	w_1\varphi_1 + w_3\varphi_3 + \kappa_0 - \kappa_1 + \kappa_2 - \kappa_3
	+ 2\rho_1 = 0.
\]
Hence the system \eqref{Eq:DS_SR_22_b} can be expressed as a system of ordinary differential equations in terms of the variables $\varphi_3,w_1,w_3$.

Let
\[
	p = \frac{w_1\varphi_3}{2t_{1,1}},\quad q = \frac{t_{1,1}w_3}{w_1},\quad
	t = t_{1,1}^2.
\]
We also set
\[\begin{split}
	&\alpha_0 = \ds\frac{1}{2}(1+\kappa_1-2\kappa_2+\kappa_3),\quad
	\alpha_1 = \ds\frac{1}{2}(-\kappa_1+\kappa_3+2\rho_1),\\
	&\alpha_2 = \kappa_0 + \kappa_2 - 2\kappa_3,\quad
	\alpha_3 = \ds\frac{1}{2}(1-2\kappa_0+\kappa_1+\kappa_3),\\
	&\alpha_4 = \ds\frac{1}{2}(-\kappa_1+\kappa_3-2\rho_1),
\end{split}\]
and
\[
	a = \alpha_0,\quad b = \alpha_3,\quad c = \alpha_4,\quad
	d = \alpha_2(\alpha_1+\alpha_2).
\]
Then we have

\begin{thm}
The system \eqref{Eq:DS_SR_22_b} with \eqref{Eq:DS_SR_22_b_BM} gives the sixth Painlev\'{e} equation.
Furthermore, $w_1$ satisfies the completely integrable Pfaffian equation
\[\begin{split}
	t(t-1)\frac{d}{dt}\log w_1 &= -(q-1)(q-t)p - \alpha_2q\\
	&\quad + \frac{1}{4}(1+2\alpha_1-2\alpha_3-4\alpha_4)t
	- \frac{1}{4}(1-2\alpha_1-4\alpha_2-2\alpha_3).
\end{split}\]
\end{thm}

\subsection{For the partition $(3,1)$}\label{Sec:System31}

The Heisenberg subalgebra $\mf{s}_{(3,1)}$ of $\mf{g}(A^{(1)}_3)$ is defined by
\[
	\mf{s}_{(3,1)} = \bigoplus_{k\in\mathbb{Z}\setminus3\mathbb{Z}}
	\mathbb{C}\Lambda_1^k\oplus\bigoplus_{k\in\mathbb{Z}\setminus\{0\}}
	\mathbb{C}z^kH_1\oplus\mathbb{C}K,
\]
with
\[
	\Lambda_1 = e_0 + e_1 + e_{2,3},\quad
	H_1 = \alpha^{\vee}_1 + 2\alpha^{\vee}_2 - \alpha^{\vee}_3.
\]
The subalgebra $\mf{s}_{(3,1)}$ admits the gradation of type $\mathbf{s}=(1,1,0,1)$ with the grade operator
\[
	\vartheta_{(3,1)}
	= 3z\left(\frac{d}{dz}+\mathrm{ad}\eta_{(3,1)}\right),\quad
	\eta_{(3,1)} = \frac{1}{3}(\alpha^{\vee}_1+\alpha^{\vee}_2+\alpha^{\vee}_3).
\]
Note that
\[
	\mf{g}_{\geq0}(1,1,0,1) = \mathbb{C}f_2\oplus\mf{b}_{+}.
\]

We now assume $t_{1,2}=1$ and $t_{1,k}=0$ $(k\geq3)$.
Then the similarity reduction \eqref{Eq:DS_SR} for $\mf{s}_{(3,1)}$ is expressed as
\begin{equation}\label{Eq:DS_SR_31_b}
	\left[\vartheta_{(3,1)}-M,\partial_{1,1}-B_{1,1}\right] = 0,
\end{equation}
with
\begin{equation}\begin{split}\label{Eq:DS_SR_31_b_BM}
	M &= \vartheta_{(3,1)}(W)W^{-1}
	+ W(\rho_1H_1+t_{1,1}\Lambda_1+2\Lambda_1^2)W^{-1},\\
	B_{1,1} &= \partial_{1,1}(W)W^{-1} + W\Lambda_1W^{-1}.
\end{split}\end{equation}

Let
\[
	W = \exp(-w_2f_2)\exp(\omega_{-1})\exp(\omega_{-2})\exp(\omega_{<-2}),
\]
where
\[\begin{split}
	\omega_{-1} &= -w_0f_0 - w_1f_1 - w_3f_3 - w_{1,2}f_{1,2}
	- w_{2,3}f_{2,3},\\
	\omega_{-2} &= -w_{0,1}f_{0,1} - w_{3,0}f_{3,0} - w_{0,1,2}f_{0,1,2} 
	- w_{1,2,3}f_{1,2,3} - w_{2,3,0}f_{2,3,0},
\end{split}\]
and $\omega_{<-2}\in\mf{g}_{<-2}(1,1,0,1)$.
Then the system \eqref{Eq:DS_SR_31_b_BM} gives explicit formulas of $M,B_{1,1}$ as follows:
\[\begin{split}
	M &= \kappa_0\alpha^{\vee}_0 + \kappa_1\alpha^{\vee}_1
	+ \kappa_2\alpha^{\vee}_2 + \kappa_3\alpha^{\vee}_3 + \varphi_0e_0
	+ (\varphi_1+w_2\varphi_{1,2})e_1\\
	&\quad + \varphi_2e_2 + (\varphi_3-w_2\varphi_{2,3})e_3
	+ \varphi_{1,2}e_{1,2} + \varphi_{2,3}e_{2,3} - 2w_2e_{3,0}
	+ 2\Lambda_1^2,\\
	B_{1,1} &= u_3K - \frac{\varphi_1-t_{1,1}}{2}\alpha^{\vee}_0
	+ \frac{\varphi_0-t_{1,1}}{2}\alpha^{\vee}_1
	+ \frac{w_2\varphi_{1,2}}{2}\alpha^{\vee}_2
	+ \frac{\varphi_{1,2}}{2}e_2 - w_2e_3 + \Lambda_1,
\end{split}\]
where
\[\begin{split}
	&\varphi_0 = 2w_1 + 2w_{2,3} + t_{1,1},\quad
	\varphi_1 = -2w_0 - 2w_{2,3} + t_{1,1},\\
	&\varphi_2 = (w_0-2w_1+t_{1,1})w_3 - 2w_{3,0},\quad \varphi_3 = 2w_{1,2},\\
	&\varphi_{1,2} = 2w_3,\quad \varphi_{2,3} = 2w_0 - 2w_1 + t_{1,1}.
\end{split}\]
Note that $\kappa_0,\ldots,\kappa_4$ are constants.
We also remark that
\[
	2w_2\varphi_2 - \varphi_3\varphi_{1,2} = 2(\kappa_2-\kappa_3-3\rho_1),\quad
	\varphi_0 + \varphi_1 + \varphi_{2,3} = 3t_{1,1}.
\]
Hence the system \eqref{Eq:DS_SR_31_b} can be expressed as a system of ordinary differential equations in terms of the variables $\varphi_0,\varphi_1,\varphi_2,\varphi_{1,2},w_2$.

Let
\[
	q_1 = -\frac{w_2\varphi_{1,2}}{\sqrt{6}},\quad
	p_1 = -\frac{2\varphi_2}{\sqrt{6}\varphi_{1,2}},\quad
	q_2 = \frac{\varphi_1}{\sqrt{6}},\quad
	p_2 = -\frac{\varphi_0}{\sqrt{6}},\quad t = -\frac{\sqrt{6}t_{1,1}}{2}.
\]
We also set
\[\begin{split}
	&\alpha_1 = \ds\frac{1}{3}(\kappa_2-\kappa_3-3\rho_1),\quad
	\alpha_2 = \ds\frac{1}{3}(\kappa_1-2\kappa_2+\kappa_3),\\
	&\alpha_3 = \ds\frac{1}{3}(1+\kappa_0-2\kappa_1+\kappa_2),\quad
	\alpha_4 = \ds\frac{1}{3}(1-2\kappa_0+\kappa_1+\kappa_3).
\end{split}\]
Then we have

\begin{thm}
The system \eqref{Eq:DS_SR_31_b} with \eqref{Eq:DS_SR_31_b_BM} gives the Painlev\'{e} system $\mathcal{H}^{A_4^{(1)}}$.
Furthermore, $\varphi_{1,2}$ satisfies the completely integrable Pfaffian equation
\[
	\frac{d}{dt}\log\varphi_{1,2} = p_1 + p_2 - \frac{2}{3}t.
\]
\end{thm}

\subsection{For the partition $(4,1)$}\label{Sec:System41}

The Heisenberg subalgebra $\mf{s}_{(4,1)}$ of $\mf{g}(A^{(1)}_4)$ is defined by
\[
	\mf{s}_{(4,1)} = \bigoplus_{k\in\mathbb{Z}\setminus4\mathbb{Z}}
	\mathbb{C}\Lambda_1^k\oplus\bigoplus_{k\in\mathbb{Z}\setminus\{0\}}
	\mathbb{C}z^kH_1\oplus\mathbb{C}K,
\]
with
\[
	\Lambda_1 = e_0 + e_1 + e_4 + e_{2,3},\quad
	H_1 = \alpha^{\vee}_1 + 2\alpha^{\vee}_2 - 2\alpha^{\vee}_3
	- \alpha^{\vee}_4.
\]
The subalgebra $\mf{s}_{(4,1)}$ admits the gradation of type $\mathbf{s}=(2,2,1,1,2)$ with the grade operator
\[
	\vartheta_{(4,1)}
	= 8\left(z\frac{d}{dz}+\mathrm{ad}\eta_{(4,1)}\right),\quad
	\eta_{(4,1)} = \frac{1}{8}
	(3\alpha^{\vee}_1+4\alpha^{\vee}_2+4\alpha^{\vee}_3+3\alpha^{\vee}_4).
\]
Note that
\[
	\mf{g}_{\geq0}(2,2,1,1,2) = \mf{b}_{+}.
\]

We now assume $t_{1,2}=1$ and $t_{1,k}=0$ $(k\geq3)$.
Then the similarity reduction \eqref{Eq:DS_SR} for $\mf{s}_{(4,1)}$ is expressed as
\begin{equation}\label{Eq:DS_SR_41_b}
	\left[\vartheta_{(4,1)}-M,\partial_{1,1}-B_{1,1}\right] = 0,
\end{equation}
with
\begin{equation}\begin{split}\label{Eq:DS_SR_41_b_BM}
	M &= \vartheta_{(4,1)}(W)W^{-1}
	+ W(\rho_1H_1+2t_{1,1}\Lambda_1+4\Lambda_1^2)W^{-1},\\
	B_{1,1} &= \partial_{1,1}(W)W^{-1} + W\Lambda_1W^{-1}.
\end{split}\end{equation}

Let
\[
	W = \exp(\omega_{-1})\exp(\omega_{-2})\exp(\omega_{-3})\exp(\omega_{-4})
	\exp(\omega_{<-4}),
\]
where
\[\begin{split}
	\omega_{-1} &= -w_2f_2 - w_3f_3,\\
	\omega_{-2} &= -w_0f_0 - w_1f_1 - w_4f_4 - w_{2,3}f_{2,3},\\
	\omega_{-3} &= -w_{1,2}f_{1,2} - w_{3,4}f_{3,4},\\
	\omega_{-4} &= -w_{0,1}f_{0,1} - w_{4,0}f_{4,0} - w_{1,2,3}f_{1,2,3}
	- w_{2,3,4}f_{2,3,4},
\end{split}\]
and $\omega_{<-4}\in\mf{g}_{<-4}(2,2,1,1,2)$.
Then the system \eqref{Eq:DS_SR_41_b_BM} gives explicit formulas of $M,B_{1,1}$ as follows:
\[\begin{split}
	M &= \kappa_0\alpha^{\vee}_0 + \kappa_1\alpha^{\vee}_1
	+ \kappa_2\alpha^{\vee}_2 + \kappa_3\alpha^{\vee}_3
	+ \kappa_4\alpha^{\vee}_4 + \varphi_0e_0 + \varphi_1e_1\\
	&\quad + \varphi_2e_2 + \varphi_3e_3 + \varphi_4e_4
	+ \varphi_{1,2}e_{1,2} + \varphi_{2,3}e_{2,3} + \varphi_{3,4}e_{3,4}
	+ 4\Lambda_1^2,\\
	B_{1,1} &= u_4K + u_0\alpha^{\vee}_0
	+ \frac{\varphi_0-2t_{1,1}}{4}\alpha^{\vee}_1 + u_2\alpha^{\vee}_2
	+ u_3\alpha^{\vee}_3 + \frac{\varphi_{1,2}}{4}e_2
	+ \frac{\varphi_{3,4}}{4}e_3 + \Lambda_1,
\end{split}\]
where
\[\begin{split}
	&\varphi_0 = 4w_1 - 4w_4 + 2t_{1,1},\quad
	\varphi_1 = -4w_0 + 2w_2w_3 - 4w_{2,3} + 2t_{1,1},\\
	&\varphi_2 = -2(2w_1-w_4-t_{1,1})w_3 - 4w_{3,4},\quad
	\varphi_3 = 2(w_1-2w_4-t_{1,1})w_2 + 4w_{1,2},\\
	&\varphi_{1,2} = 4w_3,\quad \varphi_{2,3} = -4w_1 + 4w_4 + 2t_{1,1},\quad
	\varphi_{3,4} = -4w_2,
\end{split}\]
and
\[\begin{split}
	64t_{1,1}u_0 &= (\varphi_0-4t_{1,1})(4\varphi_1+\varphi_{1,2}\varphi_{3,4})
	+ 4\varphi_2\varphi_{3,4}\\
	&\quad + 16t_{1,1}^2 + 16(\kappa_0-\kappa_1+\kappa_2-\kappa_4-2\rho_1),\\
	64t_{1,1}u_2 &= \varphi_0(4\varphi_1+\varphi_{1,2}\varphi_{3,4})
	+ 4(\varphi_2-t_{1,1}\varphi_{1,2})\varphi_{3,4}\\
	&\quad - 16t_{1,1}^2 + 16(\kappa_0-\kappa_1+\kappa_2-\kappa_4-2\rho_1),\\
	64t_{1,1}u_3 &= \varphi_0(4\varphi_1+\varphi_{1,2}\varphi_{3,4})
	+ 4\varphi_2\varphi_{3,4}\\
	&\quad - 16t_{1,1}^2 + 16(\kappa_0-\kappa_1+\kappa_2-\kappa_4-2\rho_1).
\end{split}\]
Note that $\kappa_0,\ldots,\kappa_4$ are constants.
We also remark that
\[\begin{split}
	&(\varphi_0-4t_{1,1})\varphi_{1,2}\varphi_{3,4} + 4\varphi_3\varphi_{1,2}
	+ 4\varphi_2\varphi_{3,4} = 16(-\kappa_2+\kappa_3+4\rho_1),\\
	&4\varphi_1 + 4\varphi_4 + \varphi_{1,2}\varphi_{3,4} = 16t_{1,1},\quad
	\varphi_0 + \varphi_{2,3} = 4t_{1,1}.
\end{split}\]
Hence the system \eqref{Eq:DS_SR_41_b} can be described as a system of ordinary differential equations in terms of the variables $\varphi_0,\varphi_1,\varphi_2,\varphi_{1,2},\varphi_{3,4}$.

Let
\[\begin{split}
	&q_1 = \frac{\varphi_0}{4t_{1,1}},\quad p_1 = \frac{t_{1,1}\varphi_1}{8},\\
	&q_2 = \frac{\varphi_0}{4t_{1,1}}
	+ \frac{\varphi_2}{t_{1,1}\varphi_{1,2}},\quad
	p_2 = \frac{t_{1,1}\varphi_{1,2}\varphi_{3,4}}{32},\quad
	t = -\frac{t_{1,1}^2}{2}.
\end{split}\]
We also set
\[\begin{split}
	&\alpha_1 = \frac{1}{8}(2-2\kappa_0+\kappa_1+\kappa_4),\quad
	\alpha_2 = \frac{1}{8}(2+\kappa_0-2\kappa_1+\kappa_2),\\
	&\alpha_3 = \frac{1}{8}(1+\kappa_1-2\kappa_2+\kappa_3),\quad
	\alpha_4 = \frac{1}{8}(\kappa_2-\kappa_3-4\rho_1),\\
	&\alpha_5 = \frac{1}{8}(1-\kappa_3+\kappa_4+4\rho_1).
\end{split}\]
Then we have

\begin{thm}
The system \eqref{Eq:DS_SR_41_b} with \eqref{Eq:DS_SR_41_b_BM} gives the Painlev\'{e} system $\mathcal{H}^{A_5^{(1)}}$.
Furthermore, $\varphi_{1,2}$ satisfies the completely integrable Pfaffian equation
\[
	t\frac{d}{dt}\log\varphi_{1,2} = -q_1p_1 - q_2p_2 + tq_2 - \frac{3}{4}t
	- \frac{1+2\alpha_1+2\alpha_3+2\alpha_5}{4}.
\]
\end{thm}

\appendix

\section{Lax pair}\label{Sec:Lax}

In the previous section, we have derived several Painlev\'{e} systems.
Each of them can be regarded as the compatibility condition of a Lax pair (see Remark \ref{Rem:Lax})
\[
	\frac{d\Psi}{dt} = B\Psi,\quad \vartheta_{\mathbf{n}}(\Psi) = M\Psi.
\]
In this section, we give an explicit description of $M$ and $B$ by means of a bases of $\mathfrak{sl}_{n+1}[z,z^{-1}]$.

\subsection{For the partition $(2,2)$}

The matrix $M$ is described as follows:
\[
	M = \begin{bmatrix}
		\vep_1& -\frac{2(qp+\alpha_1+\alpha_2)}{w_1}& \sqrt{t}& 0\\
		0& \vep_2& \frac{w_1(q-t)}{\sqrt{t}}& 1\\
		\sqrt{t}z& 0& \vep_3& \frac{2\sqrt{t}p}{w_1}\\
		w_1(1-q)z& z& 0& \vep_4
	\end{bmatrix},
\]
where $\vep_1,\ldots,\vep_4$ are linear conbinations of $\alpha_0,\ldots,\alpha_3$.
The matrix $B$ is expressed as follows:
\[
	B = \frac{1}{2\sqrt{t}}\begin{bmatrix}
		u_1-u_0& x_1& 1& 0\\
		0& u_2-u_1& x_2& 0\\
		z& 0& u_3-u_2& x_3\\
		x_0z& 0& 0& u_0-u_3
	\end{bmatrix}.
\]
Each component of $B$ is rational in $q,p,w_1$; see Section \ref{Sec:System22}.
The compatibility condition of this Lax pair gives the sixth Painlev\'{e} equation.

\begin{rem}
It is known that $P_{\rm{VI}}$ arises from the Lax pairs of two types, $2\times2$ matrix system {\rm\cite{IKSY}} and $8\times8$ matrix system {\rm\cite{NY3}}.
The result of this section means that we derive a new Lax pair for $P_{\rm{VI}}$.
\end{rem}

\subsection{For the partition $(3,1)$}

The matrix $M$ is described as follows:
\[
	M = \begin{bmatrix}
		\vep_1& \sqrt{6}(q_2-q_1)& \varphi_{1,2}& 2\\
		2z & \vep_2& -\frac{\sqrt{6}\varphi_{1,2}p_1}{2}& \sqrt{6}(p_2-q_2-t)\\
		\frac{2\sqrt{6}q_1}{\varphi_{1,2}}z& 0& \vep_3&
		\frac{6\{q_1(p_1+p_2-q_2-t)-\alpha_1\}}{\varphi_{1,2}}\\
		-\sqrt{6}p_2z & 2z& 0& \vep_4
	\end{bmatrix},
\]
where $\vep_1,\ldots,\vep_4$ are linear conbinations of $\alpha_0,\ldots,\alpha_3$.
The matrix $B$ is expressed as follows:
\[
	B = \frac{-2}{\sqrt{6}}\begin{bmatrix}
		u_1-u_0& 1& 0& 0\\
		0 & u_2-u_1& x_2& 1\\
		0& 0& u_3-u_2& x_3\\
		z & 0& 0& u_0-u_3
	\end{bmatrix}.
\]
Each component of $B$ is rational in $q_1,p_1,q_2,p_2,\varphi_{1,2}$; see Section \ref{Sec:System31}.
The compatibility condition of this Lax pair gives the Painlev\'{e} system $\mathcal{H}^{A_4^{(1)}}$.

Note that the system $\mathcal{H}^{A_4^{(1)}}$ also arise from the Lax pair by means of $5\times5$ matricies \cite{NY1}.

\subsection{For the partition $(4,1)$}

The matrix $M$ is described as follows:
\[\begin{split}
	M = \begin{bmatrix}
		\vep_1& \frac{8p_1}{\sqrt{-2t}}& \varphi_{1,2}& 4& 0\\
		0& \vep_2& \sqrt{-2t}\varphi_{1,2}(q_2-q_1)& 4\sqrt{-2t}(1-q_1)& 4\\
		0& 0& \vep_3& \frac{32\{(1-q_2)p_2-\alpha_4\}}{\varphi_{1,2}}&
		\frac{32p_2}{\sqrt{-2t}\varphi_{1,2}}\\
		4z& 0& 0& \vep_4& -\frac{8(p_1+p_2+t)}{\sqrt{-2t}}\\
		4\sqrt{-2t}q_1z& 4z& 0& 0& \vep_5
	\end{bmatrix},
\end{split}\]
where $\vep_1,\ldots,\vep_5$ are linear conbinations of $\alpha_0,\ldots,\alpha_4$.
The matrix $B$ is expressed as follows:
\[
	B = \frac{1}{\sqrt{-2t}}\begin{bmatrix}
		u_1-u_0& 1& 0& 0& 0\\
		0& u_2-u_1& x_2& 1& 0\\
		0& 0& u_3-u_2& x_3& 0\\
		0& 0& 0& u_4-u_3& 1\\
		z& 0& 0& 0& u_0-u_4
	\end{bmatrix}.
\]
Each component of $B$ is rational in $q_1,p_1,q_2,p_2,\varphi_{1,2}$; see Section \ref{Sec:System41}.
The compatibility condition of this Lax pair gives the Painlev\'{e} system $\mathcal{H}^{A_5^{(1)}}$.

Note that the system $\mathcal{H}^{A_5^{(1)}}$ also arise from the Lax pair by means of $6\times6$ matricies \cite{NY1}.

\subsection{For the partition $(2,2,1)$}

The matrix $M$ is described as follows:
\[
	M = \begin{bmatrix}
		0& -\frac{4\sqrt{t}\varphi_{3,4}p_2}{\varphi_3}& 
		\frac{8\sqrt{t}(q_1p_1+q_2p_2+\eta)}{\varphi_3}& 2\sqrt{t}& 0\\
		0& 0& \varphi_2& \frac{2\varphi_3(tq_2-q_1)}{t\varphi_{3,4}}& 2\\
		0& 0& 0& \frac{\varphi_3(t-q_1)}{s}& \varphi_{3,4}\\
		2\sqrt{t}z& 0& 0& 0& -\frac{4t\varphi_{3,4}p_1}{\varphi_3}\\
		\frac{2\varphi_3(q_1-q_2)}{\sqrt{t}\varphi_{3,4}}z& 2z& 0& 0& 0
	\end{bmatrix},
\]
where $\vep_1,\ldots,\vep_5$ are linear conbinations of $\alpha_0,\ldots,\alpha_4$ and
\[
	\varphi_2
	= \frac{8\{(q_2-1)(q_1p_1+q_2p_2+\eta)+\alpha_3\}}{\varphi_{3,4}}.
\]
The matrix $B$ is expressed as follows:
\[
	B = \frac{1}{2\sqrt{t}}\begin{bmatrix}
		u_1-u_0& x_1& x_{1,2}& 1& 0\\
		0& u_2-u_1& x_2& x_{2,3}& 0\\
		0& 0& u_3-u_2& x_3& 0\\
		z& 0& 0& u_4-u_3& x_4\\
		x_0z& 0& 0& 0& u_0-u_4
	\end{bmatrix}.
\]
Each component of $B$ is rational in $q_1,p_1,q_2,p_2,\varphi_3,\varphi_{3,4}$; see Section \ref{Sec:System221}.
The compatibility condition of this Lax pair gives the system \eqref{Eq:CP6} with \eqref{Eq:CP6_Ham}.

\subsection{For the partition $(3,3)$}\label{Sec:Lax33}

The matrix $M$ is described as follows:
\[
	M = \begin{bmatrix}
		\vep_1& \frac{3t^{2/3}p_1}{w_3}& \frac{1}{t^{1/3}}& 0& 0& 0\\
		0& \vep_2& \frac{w_3(t-q_1)}{t}& 1& 0& 0\\
		0& 0& \vep_3& -\frac{3(q_1p_1+q_2p_2+\eta)}{w_3}& \frac{1}{t^{1/3}}& 0\\
		0& 0& 0& \vep_4& \frac{w_3(q_2-1)}{t^{1/3}}& 1\\
		\frac{1}{t^{1/3}}z& 0& 0& 0& \vep_5& \frac{3t^{1/3}p_2}{w_3}\\
		\frac{w_3(q_1-q_2)}{t^{2/3}}z & z& 0& 0& 0& \vep_6
	\end{bmatrix},
\]
where $\vep_1,\ldots,\vep_6$ are linear conbinations of $\alpha_0,\ldots,\alpha_5$.
The matrix $B$ is expressed as follows:
\[
	B = \frac{-1}{3t^{4/3}}\begin{bmatrix}
		u_1-u_0& x_1& 1& 0& 0& 0\\
		0& u_2-u_1& x_2& 0& 0& 0\\
		0& 0& u_3-u_2& x_3& 1& 0 \\
		0& 0& 0& u_4-u_3& x_4& 0\\
		z& 0& 0& 0& u_5-u_4& x_5\\
		x_0z & 0& 0& 0& 0& u_0-u_5
	\end{bmatrix}.
\]
Each component of $B$ is rational in $q_1,p_1,q_2,p_2,w_3$; see Section \ref{Sec:System33}.
The compatibility condition of this Lax pair gives the system \eqref{Eq:CP6} with \eqref{Eq:CP6_Ham}.

\section{Affine Weyl group symmetry}\label{Sec:AffWey}

The system \eqref{Eq:CP6} with \eqref{Eq:CP6_Ham} admits affine Weyl group symmetry of type $A_5^{(1)}$.
In this section, we describe its action on the dependent variables and parameters.

Let $r_i$ $(i=0,\ldots,5)$ be birational canonical transformations defined by
\[\begin{split}
	&\alpha_0 \to -\alpha_0,\quad \alpha_1 \to \alpha_0+\alpha_1,\quad
	\alpha_5 \to \alpha_0+\alpha _5,\\
	&p_1 \to p_1 - \frac{\alpha_0}{q_1-q_2},\quad
	p_2 \to p_2 - \frac{\alpha_0}{q_2-q_1},
\end{split}\]
for $i=0$;
\[
	\alpha_0 \to \alpha_0 + \alpha_1,\quad \alpha_1 \to -\alpha_1,\quad
	\alpha_2 \to \alpha_1 + \alpha_2,\quad
	q_1 \to q_1 + \frac{\alpha_1}{p_1},
\]
for $i=1$;
\[
	\alpha_1 \to \alpha_1 + \alpha_2,\quad \alpha_2 \to -\alpha_2,\quad
	\alpha_3 \to \alpha_2 + \alpha_3,\quad
	p_1 \to p_1 - \frac{\alpha_2}{q_1-t},
\]
for $i=2$;
\[\begin{split}
	&\alpha_2 \to \alpha_2 + \alpha_3,\quad \alpha_3\to -\alpha_3,\quad
	\alpha_4\to \alpha_3 + \alpha_4,\\
	&q_1 \to q_1 + \frac{\alpha_3q_1}{q_1p_1+q_2p_2-\alpha_3+\eta},\quad
	p_1 \to p_1 - \frac{\alpha_3p_1}{q_1p_1+q_2p_2+\eta},\\
	&q_2 \to q_2 + \frac{\alpha_3q_2}{q_1p_1+q_2p_2-\alpha_3+\eta},\quad
	p_2 \to p_2 - \frac{\alpha_3p_2}{q_1p_1+q_2p_2+\eta},
\end{split}\]
for $i=3$;
\[
	\alpha_3 \to \alpha_3 + \alpha_4,\quad \alpha_4 \to -\alpha_4,\quad
	\alpha_5 \to \alpha_4 + \alpha_5,\quad
	p_2 \to p_2 - \frac{\alpha_4}{q_2-1},
\]
for $i=4$;
\[
	\alpha_0 \to \alpha_0 + \alpha_5,\quad
	\alpha_4 \to \alpha_4 + \alpha_5,\quad \alpha_5 \to -\alpha_5,\quad
	q_2 \to q_2 + \frac{\alpha_5}{p_2},
\]
for $i=5$.
Then the system \eqref{Eq:CP6} with \eqref{Eq:CP6_Ham} is invariant under the action of them.
Furthermore, a group of symmetries $\langle r_0,\ldots,r_5\rangle$ is isomorphic to the affine Weyl group of type $A_5^{(1)}$.

The group of symmetries defined above arises from the gauge transformations
\[
	r_i(\Psi) = \exp\left(\frac{\alpha_i}{\varphi_i}f_i\right)\Psi\quad
	(i=0,\ldots,5),
\]
where
\[\begin{split}
	&\varphi_0 = \frac{w_3(q_2-q_1)}{3t^{2/3}},\quad
	\varphi_1 = -\frac{t^{2/3}p_1}{w_3},\quad
	\varphi_2 = \frac{w_3\left(q_1-t\right)}{3t},\\
	&\varphi_3 = \frac{q_1p_1+q_2p_2+\eta}{w_3},\quad
	\varphi_4 = \frac{w_3(1-q_2)}{3t^{1/3}},\quad
	\varphi_5 = -\frac{t^{1/3}p_2}{w_3},
\end{split}\]
for the Lax pair of Appendix \ref{Sec:Lax33}.
Note that those transformations are derived from the following ones \cite{NY2}:
\[
	r_i(G) = G\exp(-e_i)\exp(f_i)\exp(-e_i)\quad (i=0,\ldots,5),
\]
where $G$ is an $N_{-}B_{+}$-valued function given in Section \ref{Sec:D-S}.

\section*{Acknowledgement}
The authers are grateful to Professors Kenji Kajiwara, Masatoshi Noumi, Masahiko Saito and Yasuhiko Yamada for valuable discussions and advices.


\end{document}